\documentclass[onecolumn, amsmath, amssymb, aip,groupedaddress, preprint]{revtex4-1}

\usepackage{graphicx}
\usepackage[space]{grffile}
\usepackage{dcolumn}
\usepackage{bm}
\usepackage{subfigure}
\usepackage{mathrsfs}
\usepackage{booktabs}
\usepackage{notoccite}
\usepackage{float}
\usepackage{placeins}
\usepackage{enumerate}
\usepackage{gensymb}
\usepackage{xcolor}

\begin{document}

\title{Voltage effects on the stability of Pd ensembles in Pd-Au/Au(111) surface alloys}
\author{Stephen E. Weitzner}
\email{weitzner@psu.edu}
\author{Ismaila Dabo}
\affiliation{Department of Materials Science and Engineering, Materials Research Institute, and Penn State Institutes of Energy and the Environment, \mbox{The Pennsylvania State University, University Park, PA 16802, USA } }

\begin{abstract}
The catalytic performance of multimetallic electrodes is often attributed to a beneficial combination of ligand, strain, and ensemble effects. Understanding the influence of the electrochemical environment on the stability of the alloy surface structure is thus a crucial component to the design of highly active and durable electrocatalysts. In this work, we study the effects of an applied voltage to electrocatalytic Pd-Au/Au(111) surface alloys in contact with a model continuum electrolyte. Using planewave density functional theory, two-dimensional cluster expansions are parameterized and used to simulate dilute Pd-Au surface alloys under electrochemical conditions via Metropolis Monte Carlo within an extended canonical ensemble. While Pd monomers are stable at all potentials considered, different extents of surface electrification are observed to promote the formation of Pd dimers and trimers, as well as clusters of Pd monomers. We find that the relative proportion of monomer, dimer, and trimer surface fractions is in good agreement with \emph{in situ} scanning tunneling microscopy measurements. The further development and refinement of the approaches described herein may serve as a useful aid in the development of next-generation electrocatalysts.

\end{abstract}

\maketitle

\section{Introduction}

Bimetallic nanoparticles are of fundamental interest for electrocatalysis applications as they exhibit emergent catalytic properties driven by ligand, strain, and ensemble effects that are absent in monometallic catalysts. Because of their profound impact on catalytic performance, significant attention has been devoted to understanding the roles of these promotional effects in enhancing electrocatalytic kinetics over bimetallic surfaces. It is generally believed that the observed performance enhancements arise from a beneficial modification of the local surface electronic structure, directly influencing the adsorption energies of key reaction intermediates in addition to leading to shifts in the surface Fermi level. Among these promotional mechanisms, ensemble effects are unique in that they rely on a distribution of different types of atomic ensembles along the catalyst surface to effectively activate electrochemical processes. 

The prototypical example of a bimetallic catalyst that exhibits ensemble effects are the family of palladium--gold alloy catalysts, which have been shown to be active for a number of thermochemical and electrochemical processes, such as catalyzing hydrogen evolution, low temperature carbon monoxide oxidation, hydrogen peroxide production from hydrogen and oxygen gas, vinyl acetate production, hydrocarbon hydrogenation, among other reactions.\cite{Gao2012} Model studies using both single crystals and supported bimetallic palladium--gold nanoparticles have been carried out. For example, Behm and co-workers electrodeposited palladium--gold surface alloys on single crystal gold (111) surfaces and showed via a combination of \emph{in situ} scanning tunneling microscopy, cyclic voltammetry, and \emph{in situ} Fourier transform infrared spectroscopy that carbon monoxide oxidation proceeds over palladium monomers while proton adsorption can only occur at palladium multimers containing two or more palladium atoms.\cite{maroun2001role} Goodman and co-workers found that the presence of non-adjacent but proximal palladium monomers can lead to enhanced reaction kinetics for the acetoxylation of ethylene to vinyl acetate over palladium-gold surface alloys on low index single crystal gold surfaces.\cite{chen2005promotional} Brodsky and co-workers studied supported octahedral core-shell gold-palladium nanoparticles and found that gold has a tendency to segregate to the particle surface upon potential cycling effectively diluting the palladium surface coverage, leading to an enhancement in the catalytic performance for the ethanol oxidation reaction.\cite{Brodsky2014} The extent of the gold surface segregation was observed to be highly sensitive to pH, electrolyte composition, and voltage range. Detailed electrochemical measurements of palladium-gold nanoalloys by Pizzutilo and co-workers showed that palladium may be selectively dealloyed under fuel cell operating conditions, altering the performance of surface engineered bimetallic catalysts over the course of their lifetime.\cite{Pizzutilo2017} \emph{in situ} X-ray absorption studies have also been carried out to understand the role of environmental effects on the stability of the composition and structure of palladium-gold nanocatalysts. Through extended X-ray absorption fine structure measurements, Okube and co-workers observed a strong voltage-dependence on the surface structure and proposed a mechanism through which proton adsorption at low applied potentials can draw palladium to the surface from sub-surface layers of the gold-palladium particle, generating new palladium ensembles.\cite{okube2014topologically}

To date, a number of theoretical studies have also been conducted aimed at understanding the connection between palladium-gold surfaces and their observed catalytic properties. N{\o}rskov, Behm and co-workers conducted a detailed experimental and theoretical study aimed at clarifying the adsorption of protons on palladium-gold surface alloys on palladium (111) surfaces.\cite{takehiro2014hydrogen} Employing scanning tunneling microscopy, temperature programmed desorption spectroscopy, high resolution electron energy loss spectroscopy, and semi-local density functional theory, they found that proton adsorption is most stable at compact palladium trimer three-fold hollow sites, then at palladium dimer bridge sites, and least stable at palladium monomers. They additionally found that proton adsorption at a palladium dimer--gold hollow site was more stable than at the palladium bridge site. Santos and co-workers studied near surface alloys of palladium on a gold (111) surface and found that a full sub-layer of palladium is more stable than a full monolayer on the gold (111) surface in agreement with experiment.\cite{juarez2015catalytic} They additionally found that d-electrons transfer from gold to palladium while s- and p-electrons transfer from palladium to gold with the net effect of the palladium d-band shifting up in energy towards the Fermi level. Ham and co-workers studied palladium-gold surface alloys on the palladium (111) surface and estimated the population of Pd monomers and dimers on the surface by fitting a cluster expansion Hamiltonian to density-functional results and computed averages via Monte Carlo simulations in the canonical ensemble.\cite{Ham2011} Their work showed that in the absence of adsorption effects, Pd monomers were prominent for palladium surface fractions up to 50\% for a wide range of temperatures.

While much of this work has led to profound insights into the performance and durability of bimetallic catalysts, a prominent limitation has been the ability to model bimetallic surfaces under realistic electrochemical conditions. Understanding the interplay between catalyst surface structure, performance, and stability in electrolytic environments and under applied voltage is a requirement for advancing the design of high performance electrocatalysts. In an effort to progress towards this goal, we consider a bottom-up quantum--continuum Monte Carlo approach to model the effects of solvation, applied voltage, and finite temperature on the population of palladium multimers in a palladium-gold surface alloy on a gold (111) surface. For brevity, we will refer to the system as Pd-Au/Au(111).

\section{Computational Methods}

\begin{figure}[h]
	\centering\includegraphics[width=1\columnwidth]{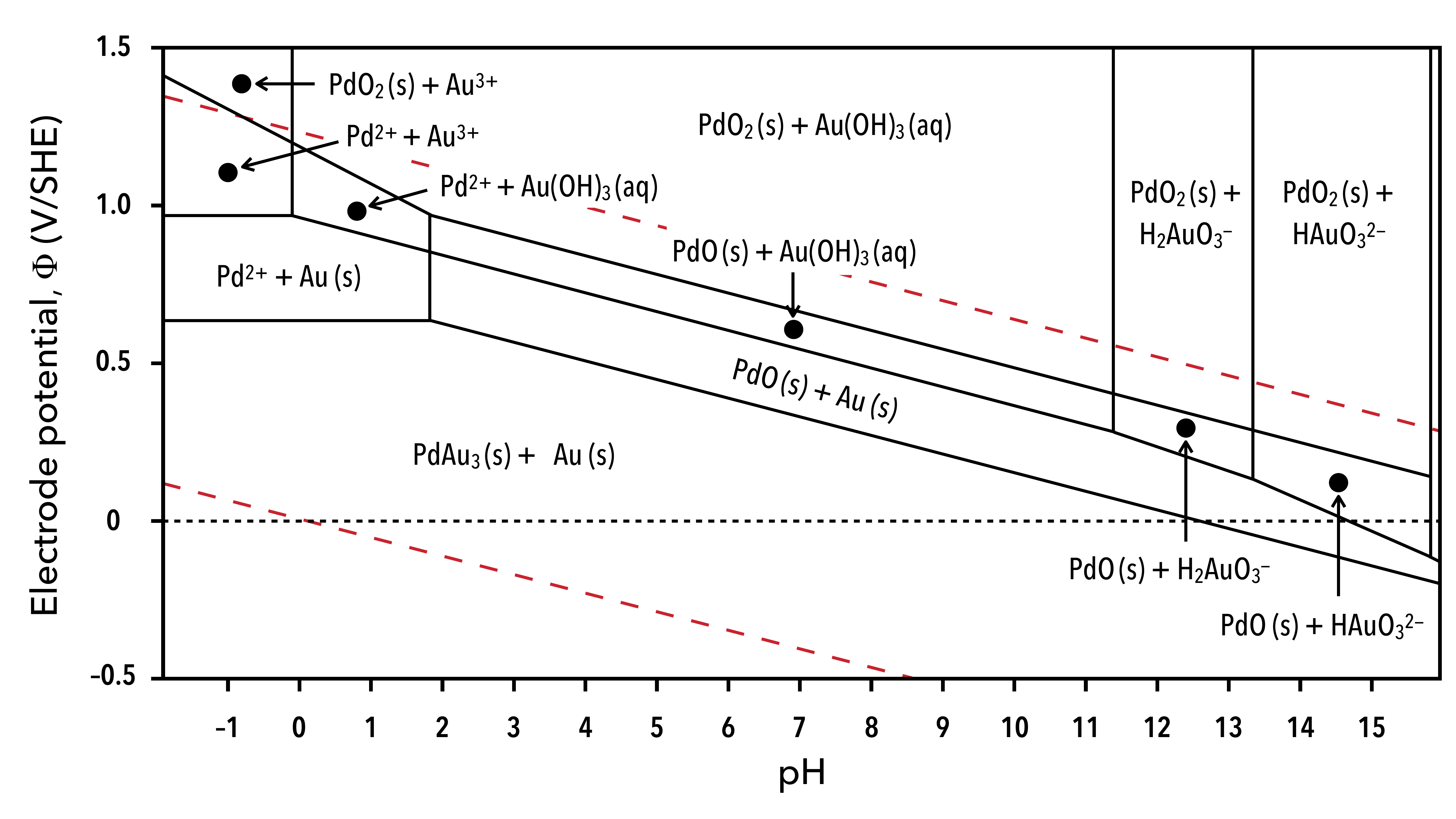}
	\caption{\small The Pourbaix diagram for a PdAu$_4$ alloy with Pd and Au solution concentrations taken to be $10^{-8}$ M.\cite{Persson2012} The lower and upper red dashed lines denote the onset of hydrogen evolution and oxygen reduction, respectively. }
	\label{fig:PdAu_Pourbaix}
\end{figure}

We model the Pd-Au/Au(111) surface alloy under applied voltage by first considering the electrochemical equilibria that exists between the bimetallic electrode and aqueous solution. As shown in Fig.~\ref{fig:PdAu_Pourbaix}, bulk gold and alloys of palladium and gold are anticipated to be thermodynamically stable at voltages between 0 V and 0.6 V vs. the standard hydrogen electrode (SHE) in strongly acidic media. 
At voltages between 0.6 V and 0.9 V, gold remains to be stable while palladium may be oxidized to form divalent Pd$^{2+}$ cations. Under high voltage conditions across a wide range of pH values, both palladium and gold may oxidize to form a variety of aqueous ions and palladium may additionally form solid PdO and PdO$_2$. In light of this diversity, we narrow the scope of our modeling efforts to study the surface alloy at a pH of 0 and between 0--0.6 V to focus solely on the effects of the applied voltage on the surface structure. It follows then, that the solution phase does not serve as a significant source of palladium and gold species. 

The bimetallic surface alloy was modeled by computing the energies of 79 unique neutrally-charged Pd-Au surface configurations in contact with a solvent using a recently developed quantum--continuum model.\cite{Weitzner2017,weitzner2017voltage,Keilbart2017} The quantum--continuum calculations were carried out using the planewave density-functional theory (DFT) code {\sc PWscf} that is part of the open-source {\sc Quantum ESPRESSO} software suite.\cite{Giannozzi2009, giannozzi2017advanced} Quantum electronic interactions were modeled using the semi-local Perdew-Burke-Ernzerhof (PBE) exchange-correlation functional. Solvent effects were described using the self-consistent continuum solvation (SCCS) model as implemented in the {\sc Environ} module that extends the {\sc PWscf} code to consider the effects of implicit liquid environments.\cite{Andreussi2012,Dupont2013} Periodic boundary artifacts on slab surfaces were additionally corrected for using the generalized electrostatic correction schemes implemented in the {\sc Environ} module.\cite{Dabo2008,Andreussi2014} The atomic cores were modeled using projector augmented wavefunction (PAW) pseudopotentials, and the wavefunction and charge density cutoffs were taken to be 50 Ry and 600 Ry, respectively, after verifying numerical convergence of forces within 5 meV/{\AA} and total energies within 50 meV per cell. The Brillouin zone of each cell was sampled with a gamma-centered $15\times 15\times 1$ Monkhorst-Pack grid, or a grid of equivalent density for larger surface cells. The electronic occupations were smoothed with 0.005 Ry of Marzari-Vanderbilt cold smearing to aid the numerical convergence of the metallic slabs.

Surfaces were modeled as symmetric slabs containing eight interior layers of pure gold and a symmetric alloy layer on the top and bottom of the slab.  Employing a similar approach to that used in Ref.~\onlinecite{Ham2011}, the surface alloy is modeled by allowing only the outermost layers to have occupational degrees of freedom. The formation enthalpy per site $\Delta H_F$ for one surface of the symmetric slabs can be computed as 
\begin{equation}
\Delta H_F(x_\text{Pd}) = \frac{1}{2 N_\text{cell} }\big[ E(N_\text{Pd})   
- N_\text{Pd} \mu^\circ_\text{Pd} 
- N_\text{Au} \mu^\circ_\text{Au}   \big]
\end{equation}
where $x_\text{Pd} = N_\text{Pd} / N_\text{cells}$ is the surface fraction of palladium, $N_\text{cell}$ is the number of surface primitive cells within the configuration, $N_\text{Au}$ is the total number of gold atoms in the slab, $E(N_\text{Pd})$ is the total quantum-continuum energy of a configuration with $N_\text{Pd}$ palladium atoms, and $\mu_\text{Pd}^\circ$ and $\mu_\text{Au}^\circ$ are the cohesive energies of bulk palladium and gold, which we have computed to be $-4.10$ eV/atom and $-3.14$ eV/atom, respectively.  We can additionally define a mixing enthalpy for the surface alloy as 
\begin{equation}
\Delta H_\text{mix} (x_\text{Pd}) =  \Delta H_F(x_\text{Pd}) - x_\text{Pd} \Delta H_F(x_\text{Pd} =1) - (1-x_\text{Pd} ) \Delta H_F(x_\text{Pd} = 0),
\end{equation}
which provides the enthalpy of each configuration relative to the pure gold (111) surface and a gold (111) surface covered with a pseudomorphic monolayer of palladium.

The formation and mixing enthalpies for the considered surfaces are shown below in Fig.~\ref{fig:form_enthalpy} and Fig.~\ref{fig:mix_enthalpy} along with the $T = 0$ K chemical potentials $\mu_\text{Pd}$ defining the equilibria amongst adjacent ground states, which are shown in red along the energy hull.
\begin{figure}[h]
	\centering\includegraphics[width=0.9\columnwidth]{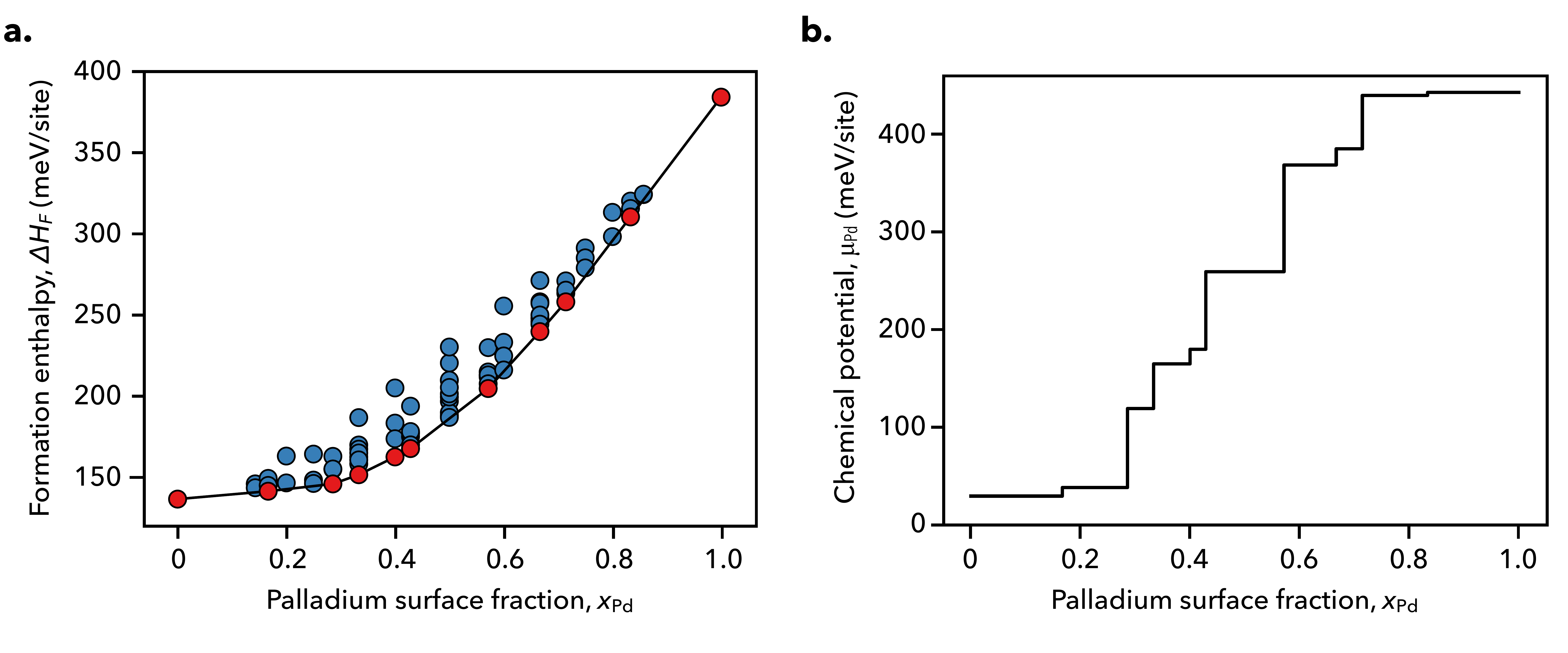}
	\caption{\small (a) The formation enthalpies of the sampled Pd-Au/Au(111) surface alloy  configurations referenced to bulk gold and bulk palladium. The ground state configurations that lie along the energy hull are shown in red. (b) The $T = 0$ K chemical potentials of the sampled Pd-Au surface alloy configurations computed from the energy hull of the formation enthalpy data.}
\label{fig:form_enthalpy}
\end{figure}
Using the formation enthalpy or mixing enthalpy results in equivalent chemical potential--composition curves, save for a linear shift in the chemical potential as a result of adopting different reference states. The ground state configurations are depicted below in Fig~\ref{fig:ground_state_configs}, where we observe that both gold and palladium prefer dispersed configurations when they exist as minority components along the surface.
\begin{figure}[h]
	\centering\includegraphics[width=0.9\columnwidth]{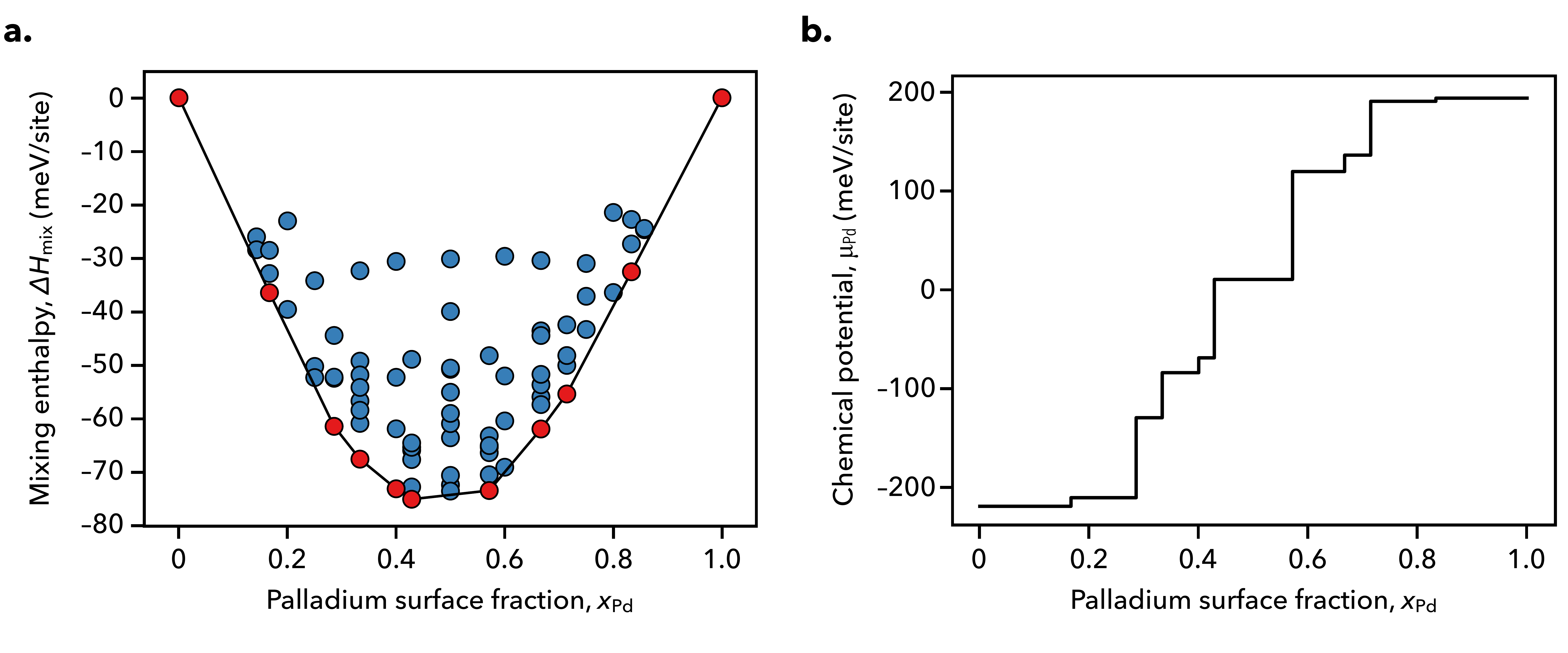}
	\caption{\small (a) The mixing enthalpies of the sampled Pd-Au/Au(111) surface alloy  configurations referenced to a pristine gold (111) surface and a gold (111) surface covered with a pseudomorphic monolayer of palladium. The ground state configurations that lie along the energy hull are shown in red. (b) The $T = 0$ K chemical potentials of the sampled Pd-Au surface alloy configurations computed from the energy hull of the mixing enthalpy data.}
\label{fig:mix_enthalpy}
\end{figure}
\begin{figure}[h]
	\centering\includegraphics[width=0.9\columnwidth]{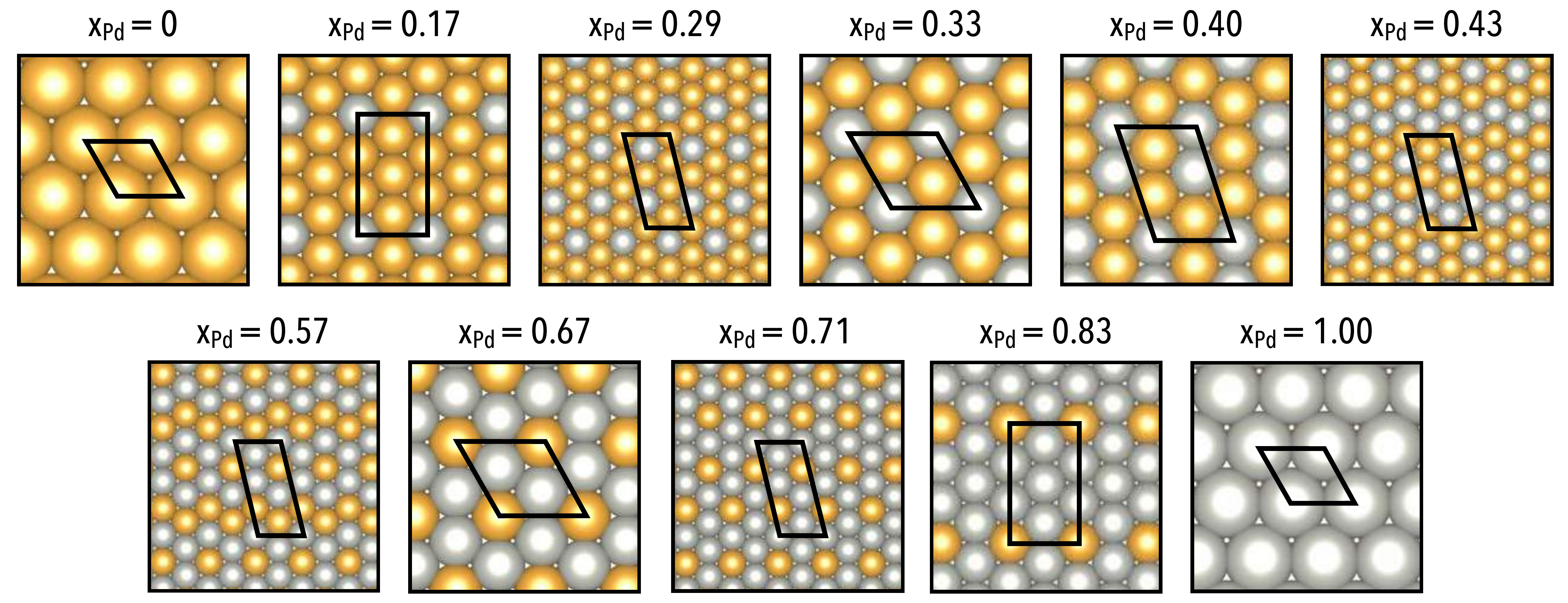}
	\caption{\small The ground state surface configurations identified in Fig.~\ref{fig:form_enthalpy} and Fig.~\ref{fig:mix_enthalpy}. In each ground state, palladium and gold is observed to mix favorably as anticipated from their negative mixing enthalpies. For surfaces with 43\% and 57\% palladium surface fractions, alternating rows of fully coordinated gold or palladium and lines of clustered gold or palladium result in the most stable configurations.}
	\label{fig:ground_state_configs}
\end{figure}

\section{The Pd-Au surface alloy under applied voltage}

As we have shown in previous studies, surface electrification effects can be considered by adding explicit charges to the supercell and inserting a planar ionic countercharge several {\AA}ngstroms from the surface within the bulk of the continuum dielectric region.\cite{Weitzner2017,weitzner2017voltage,Keilbart2017} We can then define a charge-dependent enthalpy by Taylor expanding the enthalpies obtained for neutral surfaces with respect to charge
\begin{equation}
\Delta H_\text{mix} (x_\text{Pd}, Q) = \Delta H_\text{mix} (x_\text{Pd}, Q=0) + \Phi_0(x_\text{Pd}) Q + \frac{1}{2}\frac{Q^2}{AC_0(x_\text{Pd})},
\label{eq:mix_enth_q}
\end{equation}  
where $Q$ is the total charge per site in the cell, $\Phi_0(x_\text{Pd})$ is the configuration-dependent potential of zero charge (PZC) of the alloy surface, $A$ is the surface area of one side of the slab, and $C_0(x_\text{Pd})$ is the differential capacitance of the electrode-solution interface. This capacitance term can be computed fully \emph{ab initio} for a given surface configuration by fitting Eq.~\ref{eq:mix_enth_q} to a set of enthalpies obtained for different charges and for a fixed position of the ionic countercharge.\cite{Weitzner2017,weitzner2017voltage,Keilbart2017} Alternatively, experimental capacitance values may be considered or the capacitance may be taken to be an environmental parameter and adjusted in a sensitivity analysis to approximately model the effects of the surface charge. The latter is similar in spirit to the analytical dipole corrections applied to neutral surfaces in the study of electrocatalysis.\cite{yeh2013density} This is true to the extent that a neutral surface is modeled and the effects of a perturbed interfacial electric field are approximated via an analytical correction to the configurational energies. However, unlike the dipole correction which involves an expansion in terms of the interfacial electric field, our approach  (Eq.~\ref{eq:mix_enth_q}) achieves a similar result indirectly as an expansion in terms of the surface charge. Furthermore, the expansion coefficients are identified to be more natural interfacial quantities such as the potential of zero charge and differential capacitance as opposed to the dipole moment and polarizability of surface species. We anticipate that both methods deliver equivalent accuracies and entail similar amounts of post-processing work, however a direct comparison of the methods has yet to be conducted.

\begin{figure}[h]
	\centering\includegraphics[width=0.80\columnwidth]{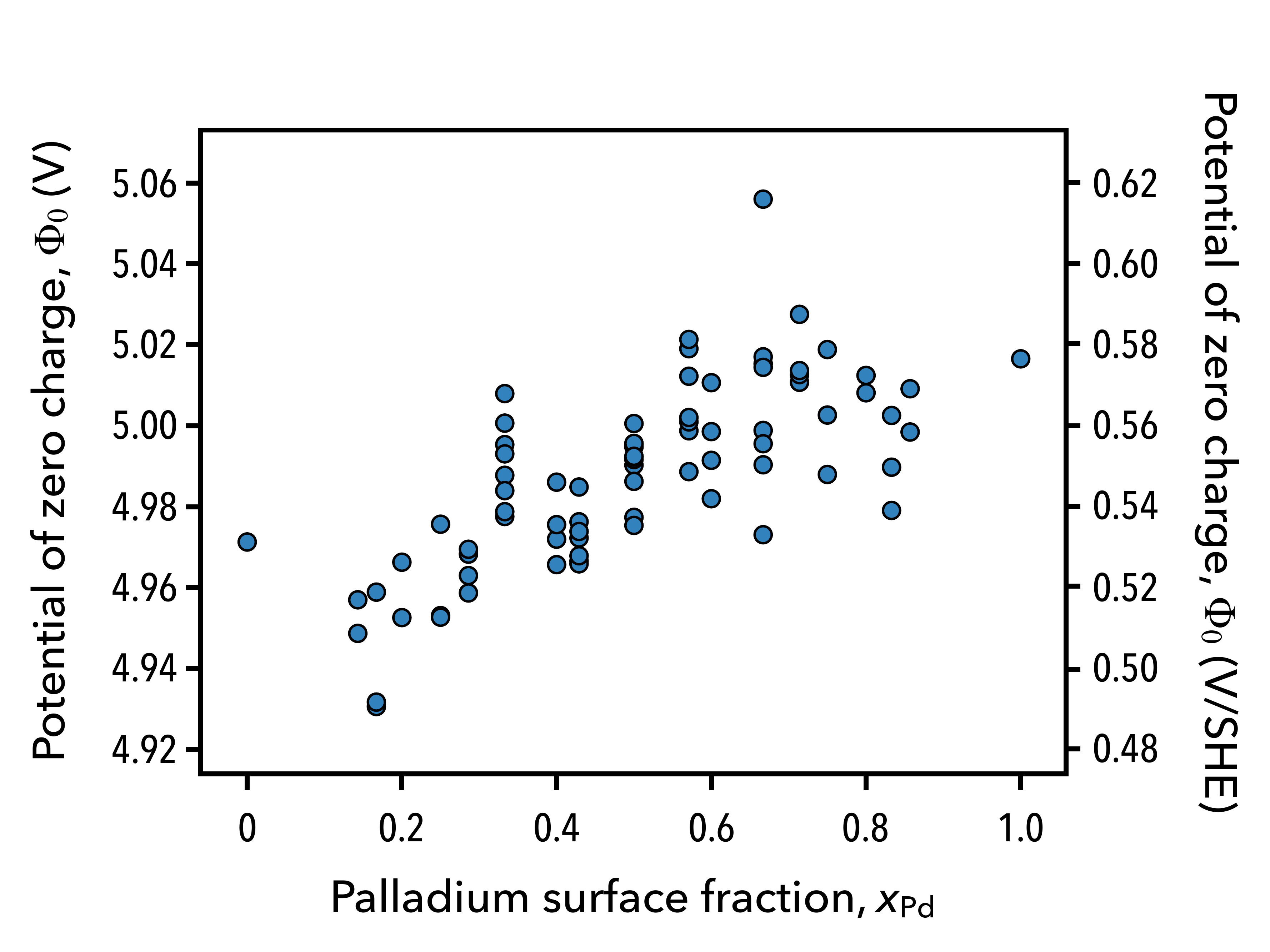}
	\caption{\small Potentials of zero charge of the sampled Pd-Au/Au(111) surface alloy configurations. Voltages are reported both on the absolute scale of the quantum--continuum calculations (left) and the standard hydrogen electrode scale (right).}
	\label{fig:au_pd_pzcs}
\end{figure}

While obtaining the enthalpy as a function of charge and composition is practical for performing quantum--continuum calculations where the charge is easily controlled, it is desirable to model the surface alloy at fixed voltages since the charge or current density is measured at a fixed potential in experiments. A voltage-dependent enthalpy or \emph{electrochemical enthalpy} may be obtained through the Legendre transform $\mathscr{F}(x_\text{Pd}, \Phi) = \Delta H_\text{mix}(x_\text{Pd}, Q) - \Phi Q$, where the voltage $\Phi$ becomes an independent potential of the system and the charge that develops on the electrode is modeled as $Q = AC_0 (\Phi - \Phi_0(x_\text{Pd}))$. In order to compute this new enthalpy, it is necessary to determine the PZC of each surface configuration, which is nothing other than the equilibrium voltage on the neutral electrode. The PZC is analogous to the work function of the neutral electrode in solution, and can be computed as $\Phi_0 = -e_0 \phi(z=\infty) - E_F$, where $e_0$ is the unsigned elementary charge, $\phi(z=\infty)$ is the electrostatic potential far from the electrode surface in the bulk of the solution, and $E_F$ is the Fermi level of the electrode. In practice, we can determine the configuration-dependent PZC by aligning the converged electrostatic potential at the edge of the supercell to zero and computing the PZC directly as $\Phi_0 = -E_F$. This provides an absolute value for the voltage relative to the bulk of the solution which must subsequently be referenced to a common standard such as the standard or reversible hydrogen electrode. This can be achieved by using Trasatti's estimate for the absolute value of the standard hydrogen electrode $\Phi_\text{SHE} = 4.44$ V, or by aligning the set of PZC values to an experimentally determined PZC so that the computed PZC of the neutral pristine surface matches the experimental one.\cite{trasatti1986absolute} In this work, we consider a non-reconstructed Au(111) surface for which PZC data is scarce. We therefore reference the voltages in these simulations by subtracting $\Phi_\text{SHE}$, as shown in Fig.~\ref{fig:au_pd_pzcs}.
Here, we observe that the PZC of the gold (111) surface increases by 50 mV after replacing the top surface layer with a full palladium monolayer. An increase in the PZC after metal electrodeposition is not very surprising since the palladium (111) surface has a work function of 5.6 eV compared to a work function of 5.3 eV for the gold (111) surface, and a similar trend is to be expected for the PZCs.\cite{michaelson1977work} It has also been shown previously that thin electrodeposited metal films often exhibit PZCs between that of the substrate surface and a bulk surface of the depositing metal.\cite{el2006potential} 

\begin{figure}[h]
	\centering\includegraphics[width=1\columnwidth]{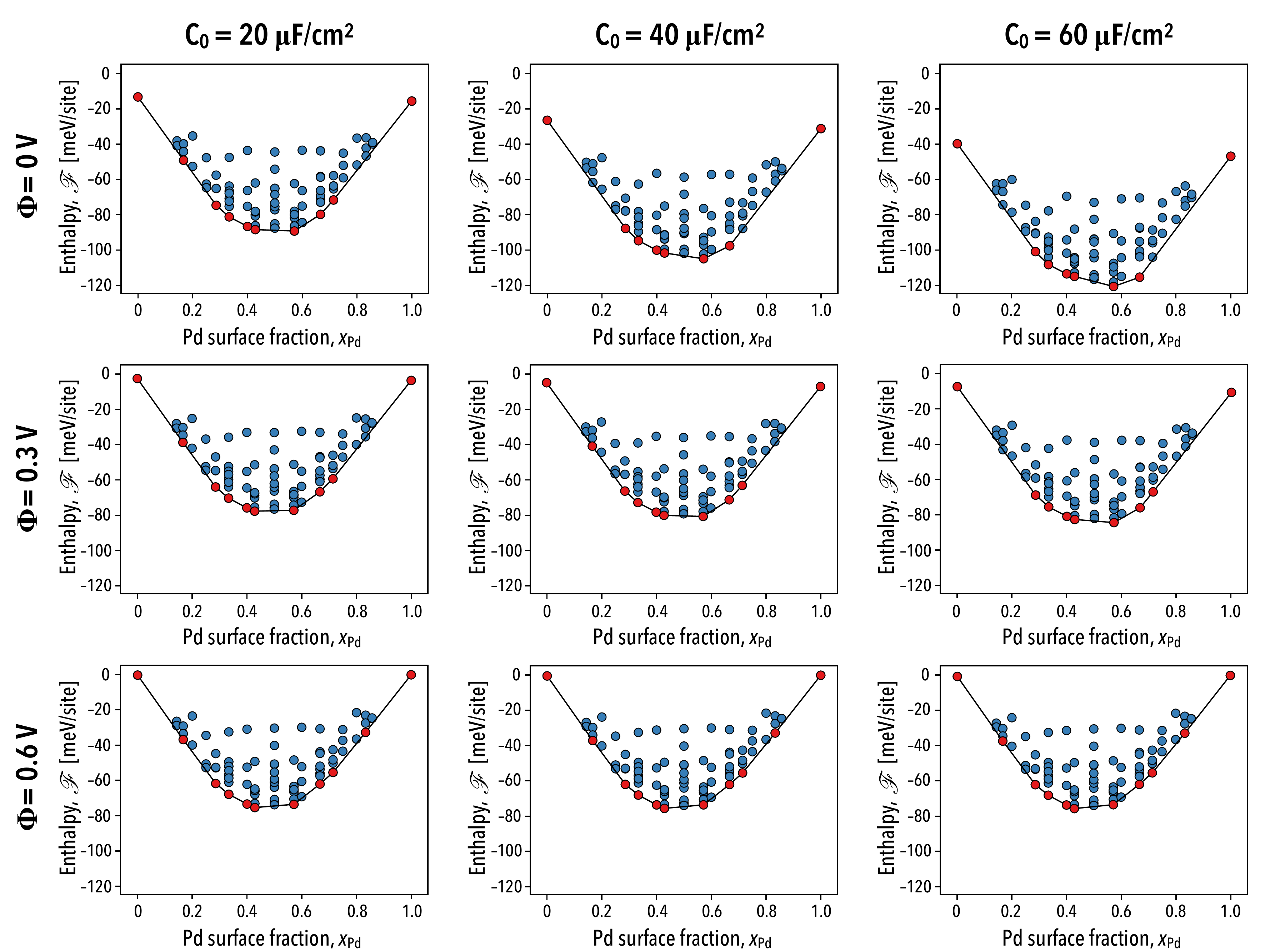}
	\caption{\small The electrochemical enthalpies at several values of the applied voltage and differential capacitance. At a voltage of 0.6 V, the surface charge tends towards zero as the PZC of the Pd-Au/Au(111) surface varies approximately between 0.50 and 0.60 V. The mixing enthalpies thus approach the limiting case of the neutral surface. At lower voltages, a negative charge develops leading to a strong perturbation of the ground states. The effects of this surface charging process is most evident when the differential capacitance is increased, since this leads to a concomitant increase in the magnitude of the surface charge. A nonuniform shift in the mixing enthalpy curve occurs, where the configurations with palladium surface fractions near 60\% are stabilized more strongly than more dilute configurations due to the differences in their PZCs.}
	\label{fig:electrochem_mixing_enthalpies}
\end{figure}

In Fig.~\ref{fig:electrochem_mixing_enthalpies}, we show how the applied voltage affects the mixing enthalpies of the Pd-Au/Au(111) surface alloy. 
At low potentials, surface electrification effects become prominent and we observe that the ground state at $x_\text{Pd} = 0.57$ becomes increasingly stabilized with an increasingly negative surface charge. At low voltages, the neutral ground state configurations at $x_\text{Pd} = 0.25$ and $x_\text{Pd} = 0.83$ move away from the energy hull. This indicates two possibilities: that either new two-phase regions in these composition ranges appear under electrochemical conditions, or that new ground state configurations with surface cells larger than those considered to construct the training set may exist. Further study will be required to clarify this observation.

\section{Cluster expansion fitting}

\begin{figure}[h]
	\centering\includegraphics[width=1\columnwidth]{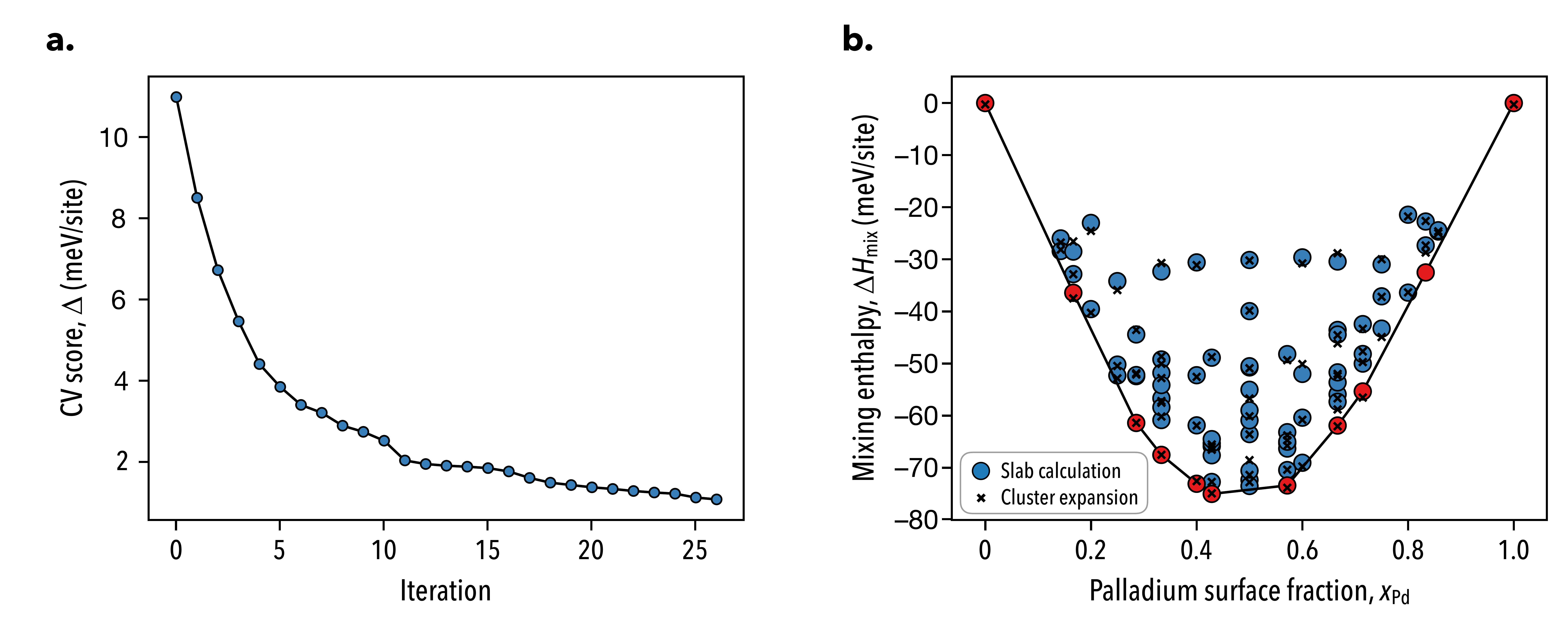}
	\caption{\small Results of the cluster expansion fitting process showing a) the convergence of the LOOCV score for the neutral Pd-Au/Au(111) surface, and b) the predicted cluster expansion enthalpies for the training set using the optimized set of clusters. The cluster expansion provides accurate enthalpy estimates for lower energy configurations with dilute palladium surface fractions.}
	\label{fig:LOOCV_convergence}
\end{figure}

Following the procedure outlined in Ref.~\onlinecite{weitzner2017voltage}, two-dimensional cluster expansions were fitted to electrochemical enthalpies computed at voltages of 0, 0.3 and 0.6 V vs. SHE and for differential capacitance values of 20, 40, and 60 $\mu$F/cm$^2$.\cite{el2002potentials} We used an in house code to perform the cluster expansion fitting that implements the steepest descent approach described in Ref.~\onlinecite{Herder2015}. We enforced the restriction that all proposed expansions contained the empty, point, and nearest-neighbor pair clusters to ensure that local interactions were adequately described in the expansion.  The remaining clusters in the final expansion were included by minimizing a leave-one-out cross-validation (LOOCV) score $\Delta$ for the entire training set using the steepest descent approach. Briefly, the score is computed as 
\begin{equation}
\Delta = \left(  \frac{1}{k} \sum_k (\mathscr{F}(\{ \sigma_i \}) - \hat{\mathscr{F}}(\{ \sigma_i \})^2 \right)^{\frac{1}{2}},
\end{equation}
where the mean square error of the configurational energy is computed for which the cluster expansion estimate $\hat{\mathscr{F}}(\{ \sigma_i \})$ is computed with a set of ECIs obtained from a fit excluding the current configuration in the training set being considered. We allowed clusters that contained up to four vertices and with maximal diameters of up to nine nearest neighbors to be included in the cluster search space. In panel a of Fig.~\ref{fig:LOOCV_convergence}, we show the convergence of the LOOCV score as a function of the number of iterations of the steepest descent algorithm. 
Typically LOOCV scores are on the order of tens of meV/site, however we obtain converged results on the order 1 meV/site due to the small magnitude of the surface alloy mixing enthalpy. In panel b of Fig.~\ref{fig:LOOCV_convergence}, we show the cluster expansion estimates for the mixing enthalpies. Overall, the cluster expansion leads to a good fit of the training set, however higher energy alloys are predicted less accurately, as well as some high palladium content surface alloys. Neither of these pose issues to the present analysis since we perform simulations at room temperature where high energy configurations are infrequently sampled and we furthermore restrict our analysis to low-palladium content surfaces.
 
\section{Fixed-voltage canonical Monte Carlo simulations}

\begin{figure}[h]
	\centering\includegraphics[width=1\columnwidth]{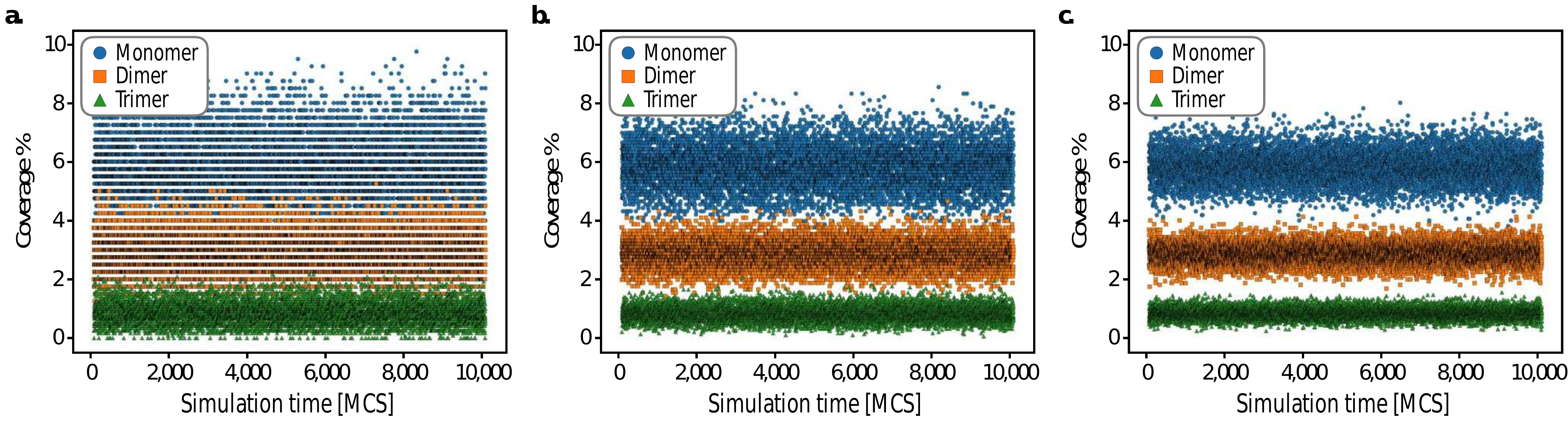}
	\caption{\small Convergence of monomer, dimer, and trimer distributions for simulation cell sizes of a) $20\times 20$, b) $30\times 30$, and c) $40\times 40$ primitive cells at a fixed capacitance of 60 $\mu$F/cm$^2$ and at a voltage of $\Phi = 0.3$ V/SHE.}
	\label{fig:finite_size_effects}
\end{figure}

The Pd-Au/Au(111) surface alloy was studied at a set of fixed voltages within an extended canonical ensemble $(N_\text{Au}, N_\text{Pd},V,T,\Phi)$ via Metropolis Monte Carlo. The associated Boltzmann probability in this ensemble takes the form
\begin{equation}
\mathcal{P}_i = \frac{1}{\mathcal{Z}} \exp\left[ -\beta N_\text{cell}\Delta \mathscr{F}(\Phi) \right],
\end{equation}
where $\mathcal{Z}$ is the partition function, $\beta = \frac{1}{k_B T}$, $N_\text{cell}$ is the number of primitive surface cells in the system, and $\Delta \mathscr{F}(\Phi) = \Delta H_\text{mix}(\{\sigma_i\}) - \Phi \Delta Q(\{\sigma_i\}, \Phi) $ is the difference in electrochemical enthalpy between subsequently generated states in the simulation. New states are proposed via spin-exchange trial moves, which consist of randomly selecting a pair of opposite spins on the lattice and exchanging them, thereby preserving the overall composition of the system while ergodically exploring the configurational space.\cite{Newman1999} Trajectory data was analyzed for temporal correlations in the monomer, dimer, and trimer coverages and were found to be fully decorrelated within one Monte Carlo sweep (MCS). Samples were thus collected after each sweep of the lattice, where one sweep consists of performing a number of random Monte Carlo moves equal to the number of sites in the lattice. Finite size effects were additionally tested for as shown in Fig.~\ref{fig:finite_size_effects}, and it was found that a cell size of $40\times 40$ primitive cells led to a good balance of precision and computational cost.
Simulations were allowed to equilibrate for 100 MCS prior to computing average multimer coverages over 10,000 MCS. This was sufficient to obtain averages for the palladium ensemble coverages converged to within a precision of $10^{-4}$.

\section{Results and discussion}
In order to assess the predictive accuracy of the quantum--continuum model and the sensitivity of the palladium multimer coverage distributions to solvation and surface electrification effects, we make a comparison with coverage measurements performed by Behm and co-workers via \emph{in situ} scanning tunneling microscopy.\cite{maroun2001role} To facilitate the comparison, canonical Monte Carlo simulations were performed for palladium surface fractions of $x_\text{Pd} = 0.07$ and  $x_\text{Pd} = 0.15$ at voltages of 0, 0.3, and 0.6 V vs. SHE. We additionally consider differential capacitance values of 20, 40, and 60 $\mu$F/cm$^2$ in accordance with capacitance measurements made by Kolb and co-workers for a palladium monolayer-covered gold (111) surface in a 10 mM NaF electrolyte.\cite{el2002potentials} 

\begin{figure}[h]
	\centering\includegraphics[width=1\columnwidth]{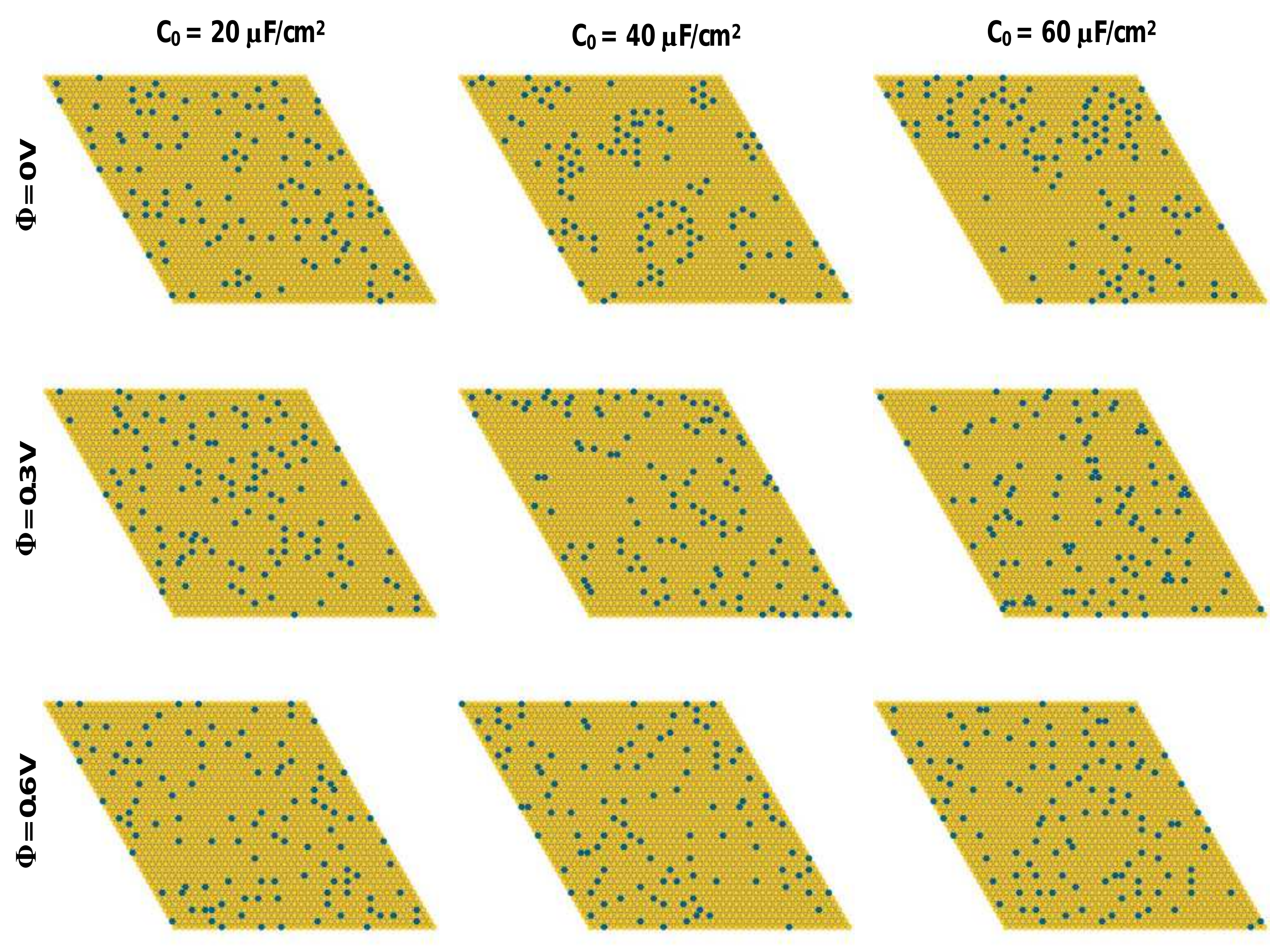}
	\caption{\small Snapshots of the palladium--gold surface alloy for a palladium surface fraction of $x_\text{Pd} = 0.07$ for different voltages and differential capacitances. The surface palladium atoms are shown in blue. Monomer clustering is evident at 0 V, while dimer and trimer formation can also be clearly seen at 0.3 V. At higher potentials, palladium adopts a more dispersed state along the surface.}
	\label{fig:x_pd_07_snapshots}
\end{figure}

\begin{figure}[h]
	\centering\includegraphics[width=1\columnwidth]{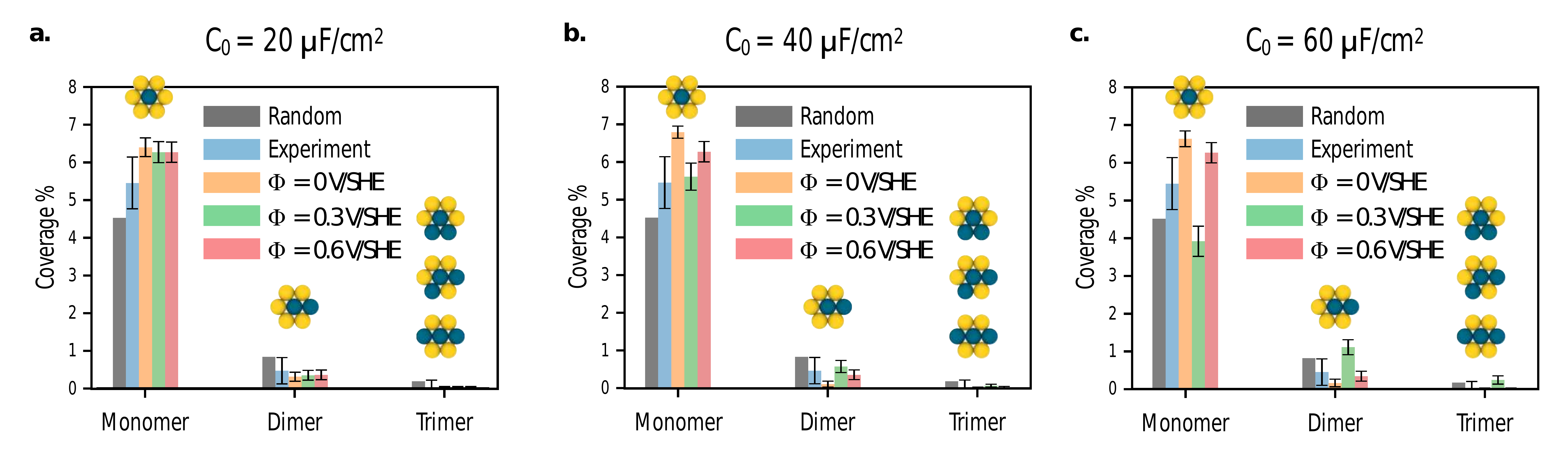}
	\caption{\small Average palladium ensemble coverage for a palladium surface fraction of $x_\text{Pd} = 0.07$ under applied applied voltage for differential capacitance values of a) 20 $\mu$F/cm$^2$, b) 30 $\mu$F/cm$^2$, c) 60 $\mu$F/cm$^2$. Error bars for the Monte Carlo data are the standard deviation of each coverage distribution. Experimental data and random alloy data were obtained from Ref.~\onlinecite{maroun2001role}.}
	\label{fig:ave_multimer_007}
\end{figure}

In Fig.~\ref{fig:x_pd_07_snapshots}, we show several snapshots of the Pd-Au/Au(111) surface alloy with a composition of $x_\text{Pd} = 0.07$ obtained for simulations run under different electrochemical conditions. 
We observe that palladium monomers are the dominant type of multimer for all cases, and that systems with lower degrees of surface electrification achieved with either higher voltage or lower differential capacitance tend to adopt more dispersed configurations. For surfaces with higher degrees of surface electrification achieved via higher differential capacitance values or with lower voltages, we find that palladium tends to cluster along the surface. Interestingly, we find two particular cases of clustering where higher order dimer and trimer multimers appear to be stabilized at intermediate voltages, while at low voltages palladium surface atoms cluster to form locally ordered regions with palladium monomers situated at second nearest neighbor positions. In Fig.~\ref{fig:ave_multimer_007}, we make a quantitative comparison of the Monte Carlo multimer coverage estimates with the experimental results reported in Ref.~\onlinecite{maroun2001role}. 
For each value of the differential capacitance considered, we obtain close agreement with the experimentally measured multimer coverages. We observe the general trend that the monomer coverage is highest at low voltages, decreases at intermediate voltages with an increased stabilization of dimers, and then increases again at higher voltages. 

\begin{figure}[h]
	\centering\includegraphics[width=1\columnwidth]{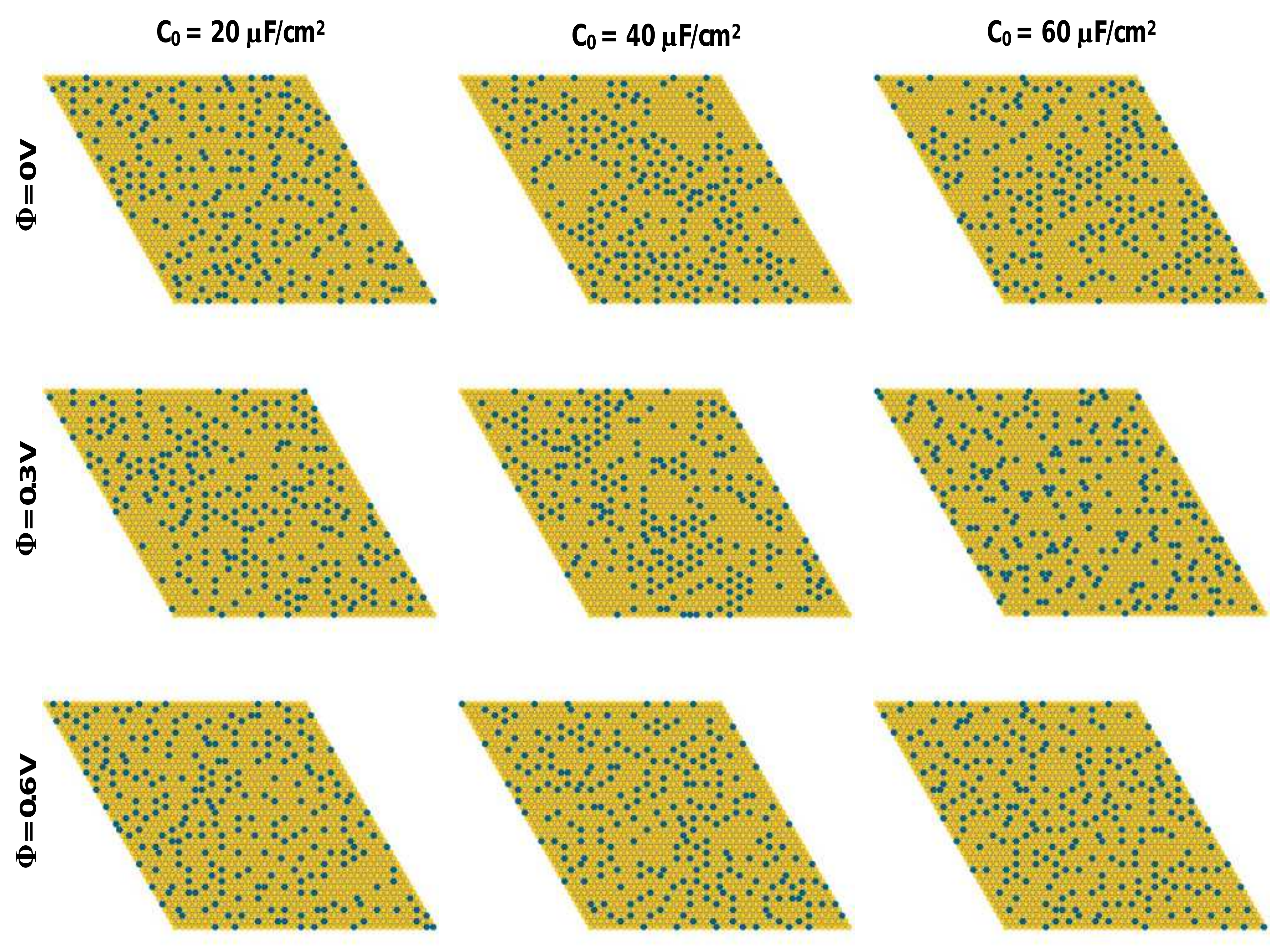}
	\caption{\small Snapshots of the palladium--gold surface alloy for a palladium surface fraction of $x_\text{Pd} = 0.15$ for different voltages and differential capacitances. The surface palladium atoms are shown in blue. Monomer clustering is evident at 0 V and 0.3 V, while dimer and trimer formation can also be clearly seen at 0.3 V. Similar to the more dilute case, palladium adopts a more dispersed state along the surface at higher potentials.}
\label{fig:x_pd_15_snapshots}
\end{figure}

\begin{figure}[h]
	\centering\includegraphics[width=1\columnwidth]{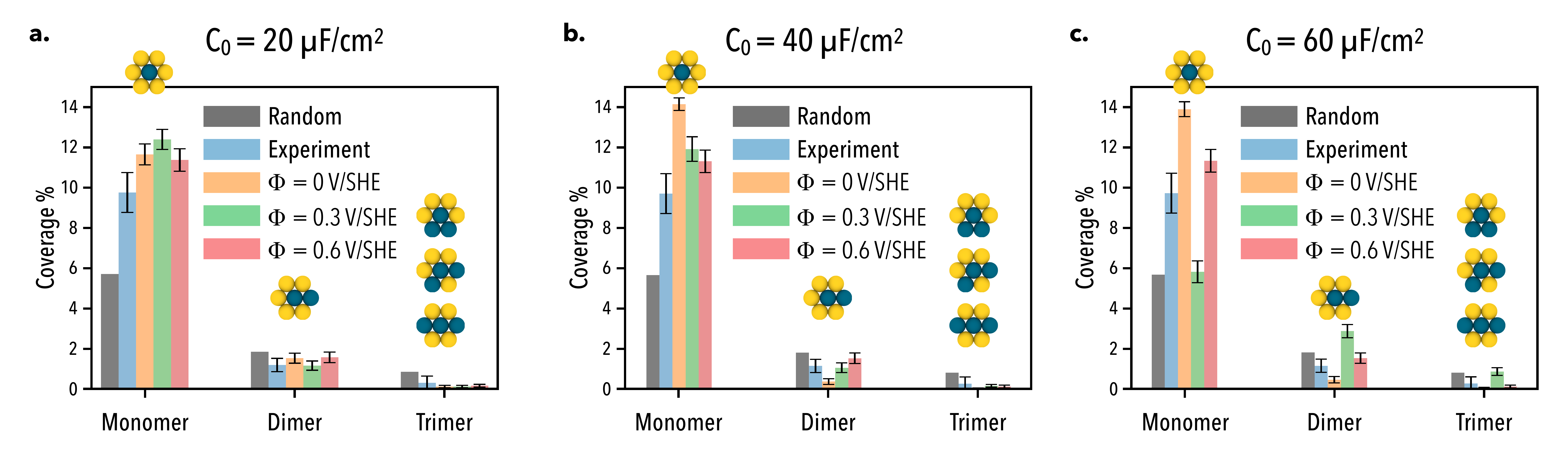}
	\caption{\small Average palladium ensemble coverage for a palladium surface fraction of $x_\text{Pd} = 0.15$ under applied applied voltage for differential capacitance values of a) 20 $\mu$F/cm$^2$, b) 30 $\mu$F/cm$^2$, c) 60 $\mu$F/cm$^2$. Error bars for the Monte Carlo data are the standard deviation of each coverage distribution. Experimental data and random alloy data were obtained from Ref.~\onlinecite{maroun2001role}.}
	\label{fig:ave_multimer_015}
\end{figure}

Similar behavior is observed for the Pd-Au/Au(111) surface alloy with a palladium surface fraction of $x_\text{Pd} = 0.15$. We show in Fig.~\ref{fig:x_pd_15_snapshots} several snapshots of the surface alloy simulated under the same set of electrochemical conditions as the dilute surface considered previously.
Like the dilute composition, we find that palladium monomers tend to be the dominant multimer under most of the considered electrochemical conditions except for large values of the differential capacitance at intermediate voltages. In this case we see a pronounced stabilization of dimers and trimers that appear to be uniformly distributed over the surface. In addition to this, we observe that palladium tends to exhibit the same type of ordering identified in the dilute case at low potentials, where palladium monomers are locally clustered sitting at second nearest neighbor positions from one another. In Fig.~\ref{fig:ave_multimer_015}, we compare the Monte Carlo multimer coverage estimates to the \emph{in situ} scanning tunneling microscopy results reported by Behm and co-workers.
We again find our results to be in good agreement with experiment, however in this case we observe a stronger response to the applied voltage and differential capacitance as compared to the dilute surface alloy. For almost all sets of electrochemical conditions, we predict a slightly higher monomer coverage for surface alloys simulated with low differential capacitance values. As the differential capacitance is increased, we find an enhanced stabilization of monomers at low voltages and an enhanced stabilization of dimers and trimers at intermediate voltages. 

It is worthwhile to note that for both of the surface alloy compositions considered in this analysis, the closest results with experiment were found for surfaces considered at $\Phi = 0.6$ V vs. SHE, close to the potential at which the surface alloys were electrodeposited.\cite{maroun2001role} While this result is promising, it is important to point out that the adsorption of both protons and anions such as sulfate or bisulfate are known to occur and may play an important role in determining the composition and therefore the distribution of multimers along the surface.\cite{okube2014topologically, maroun2001role} This is especially true for palladium--gold nanoparticles where surface segregation effects are known to occur; however, the influence of these co-adsorbates is less clear for model surface alloys on single crystal surfaces where the active components are restricted to the topmost surface layer.\cite{Brodsky2014} It is additionally promising to see that accounting for the capacitive nature of the interface can lead to a measurable change in the equilibrium distribution of palladium multimers along the surface, indicating that the type and distribution of active sites along the surface exhibits a voltage-dependence that is independent of co-adsorption effects. 

\section{Summary}
In this work, a quantum--continuum model was applied to study the effects of solvation and surface electrification on the equilibrium distribution of palladium multimers in a palladium--gold surface alloy on the gold (111) surface. Electrochemical enthalpies obtained with the quantum--continuum model were used to fit two-dimensional cluster expansions of the surface alloy for different sets of voltages and differential capacitances, defining several different electrochemical environments. Metropolis Monte Carlo simulations were performed in the canonical ensemble for fixed voltages using non-local spin-exchange moves. Close agreement with experimentally measured palladium multimer coverages was found for each case considered. We found that at voltages near 0 V vs. SHE, palladium monomers are predicted to be stable and tend to adopt locally ordered structures with neighboring palladium atoms occupying second-nearest neighbor positions. At voltages near 0.3 V vs SHE, we found that palladium dimers and trimers are stable and homogeneously distributed along the surface when the differential capacitance approaches 60 $\mu$F/cm$^2$, but adopts similar low voltage configurations for lower differential capacitances. At voltages near 0.6 V vs SHE, palladium is observed to exist primarily as monomers along the surface. These results suggest that applied voltages can provide a driving force for the ordering or clustering of catalytically active multimers within surface alloys, altering the distribution and variety of active sites along the catalyst surface that are available for electrocatalysis under different electrochemical conditions. This work provides a new perspective and direction for modeling the durability of electrocatalytic alloys in electrochemical environments.

\begin{acknowledgements}
This work was supported by the U.S. Department of Energy, Office of Science, Basic Energy Sciences, CPIMS Program, under Award \# DE-SC0018646. Computations for this research were performed on the Pennsylvania State University's Institute for CyberScience Advanced CyberInfrastructure (ICS-ACI).
\end{acknowledgements}


%

\end{document}